\documentclass[journal]{IEEEtran}
\usepackage{amsmath}
\usepackage{amssymb}
\usepackage{amsthm}
\usepackage{cite}
\usepackage{color}
\usepackage{xcolor}
\usepackage{caption}
\usepackage{epsfig,latexsym}
\usepackage{float}
\usepackage{fancyhdr}
\usepackage{graphicx}
\usepackage{indentfirst}
\usepackage{mdwmath}
\usepackage{mdwtab}
\usepackage{subfigure}
\usepackage{setspace}
\usepackage{times}
\usepackage{url}
\usepackage{multirow}
\usepackage{algorithm}
\usepackage{algorithmic}
\usepackage{verbatim}
\usepackage{algorithm,algorithmic}
\usepackage{mathrsfs}

\newtheorem{remark}{Remark}
\newtheorem{theorem}{Theorem}

\newtheorem{lemma}{Lemma}

\newtheorem{corollary}{Corollary}

\begin{document}

\title{\huge{Pinching-Antenna Systems (PASS)-based Indoor Positioning}}

\author{Yaoyu Zhang, Xin Sun, Jun Wang, Tianwei Hou,~\IEEEmembership{Member,~IEEE}, Anna Li,~\IEEEmembership{Member,~IEEE},\\ 
Yuanwei Liu,~\IEEEmembership{Fellow,~IEEE}, and Arumugam Nallanathan,~\IEEEmembership{Fellow,~IEEE}
       
\thanks{Yaoyu Zhang,  Xin Sun, Jun Wang and Tianwei Hou are with the School of Electronic and Information Engineering, Beijing Jiaotong University, Beijing 100044, China (e-mail: 24110070@bjtu.edu.cn; xsun@bjtu.edu.cn; wangjun1@bjtu.edu.cn; twhou@bjtu.edu.cn). }  
\thanks{Anna Li is with the School of Computing and Communications, Lancaster University, Lancaster LA1 4WA, U.K. (e-mail: a.li16@lancaster.ac.uk). }
\thanks{Yuanwei Liu is with the Department of Electrical and Electronic Engineering, The University of Hong Kong, Hong Kong (e-mail: yuanwei@hku.hk). } 
\thanks{Arumugam Nallanathan is with the School of Electronic Engineering and Computer Science, Queen Mary University of London, London E1 4NS, U.K., and also with the Department of Electronic Engineering, Kyung Hee University, Yongin-si, Gyeonggi-do 17104, Korea (e-mail: a.nallanathan@qmul.ac.uk).}}

\maketitle
\vspace{-0.1in}
\begin{abstract}
Pinching antenna (PA), a flexible waveguide integrated with dielectric particles, intelligently reconstructs line-of-sight channels. Utilizing its geometric deterministic model and meter-level reconstruction, PA systems (PASS) are applied to uplink indoor positioning. In this paper, the uplink positioning system model for PASS is firstly proposed. A PASS-based received signal strength indication (RSSI) method is proposed to measure the distance from the users to each PA, which is efficient and suitable for PASS. PASS-based weighted least squares (WLS) algorithm is designed to calculate the two-dimensional coordinates of the users. Several critical observations can be drawn from our results: i) More PAs on the waveguide improves the positioning accuracy and robustness. ii) When the number of PAs exceeds a certain threshold, the performance gain becomes marginal. iii) User locations between and near PAs yield superior positioning accuracy.
\end{abstract}


\section{Introduction}

Flexible antennas are gaining increasing attention as a promising solution to meet the evolving demands of modern wireless communication systems. The technologies such as reconfigurable intelligent surfaces~\cite{RIS}, dynamic meta-surface antennas~\cite{dynamic}, fluid antennas~\cite{fluid_antenna}, and movable antennas~\cite{movable_antenna} enable real-time reconfiguration of radiation patterns, offering  improved performance in dynamic environments. However, the physical reconfiguration range of such flexible antenna systems is fundamentally constrained by operating wavelength, limiting the spatial adaptability to meter-scale adjustments of antenna systems in millimeter-wave bands. In contrast, the pinching-antenna systems (PASS) break the constraint by enabling continuous adjustments over meter-scale distances through dielectric waveguides, achieving unprecedented flexibility in electromagnetic field control~\cite{PA0}. 

Pioneered by Suzuki et al., PASS exploits mechanical deformation of waveguide structures to dynamically control electromagnetic fields without electronic phase shifters, enabling macroscopic port mobility unattainable by conventional flexible antennas~\cite{PA2022}. Subsequent research has focused exclusively on the communication performance of PASS. Hou et al. has comprehensively evaluated the performance gains of PASS in uplink transmissions~\cite{PASS_UL}. Xu et al. has optimized the locations of the pinching antennas to maximize the downlink transmission rate~\cite{PASS_rate}, while Xiao et al. explored channel estimation for PASS~\cite{PASS_channel}. Despite these advances, existing studies are still limited to communication-centric metrics such as communication rate and channel estimation, ignoring the intrinsic advantages of PASS in spatial localization.

While indoor positioning systems increasingly leverage reconfigurable hardware, PASS remains unexplored for localization. State-of-the-art systems employ reconfigurable intelligent surface (RIS) or phased arrays to improve angle-of-arrival estimation, which suffer from hardware complexity and sensitivity to orientation~\cite{RIS_positioning},~\cite{phased_array_positioning}. PASS circumvents these issues through its unified waveguide architecture. Nevertheless, no existing study has investigated the potential of PASS for high-accurate localization, creating a critical knowledge gap at the intersection of antenna design and positioning. Considering the geometrically deterministic channel model and meter-scale reconfigurability of PASS, it is ideal for indoor positioning.

Motivated by this gap, we propose the first PASS-based indoor positioning framework. The following is a brief summary of our main contributions in this letter:
\begin{itemize}
  \item We propose a novel PASS-based indoor positioning model incorporating dielectric waveguide loss, which is more consistent with the actual waveguide transmission and has significant impact on positioning accuracy.  
  \item We measure the distance from each PA to the user by utilizing the received signal strength indication (RSSI) method. The PASS-based weighted least squares (WLS) positioning algorithm is designed to calculate the location of user. The results show that increasing the number of PAs can improve positioning accuracy. 
  \item The simulation results demonstrate that: 1) Increasing the number of PAs on the waveguide reduces the positioning error and enhances the robustness of user positioning. 2) when the number of PAs exceeds 7, the performance gain decreases with increasing number of PAs. 3) When the user is located between the PAs and in close proximity to them, the positioning performance is improved.
\end{itemize}

\section{System Model}

\subsection{Antenna and Channel Model}

\begin{figure}[t!]
\centering
\includegraphics[width =3.5in]{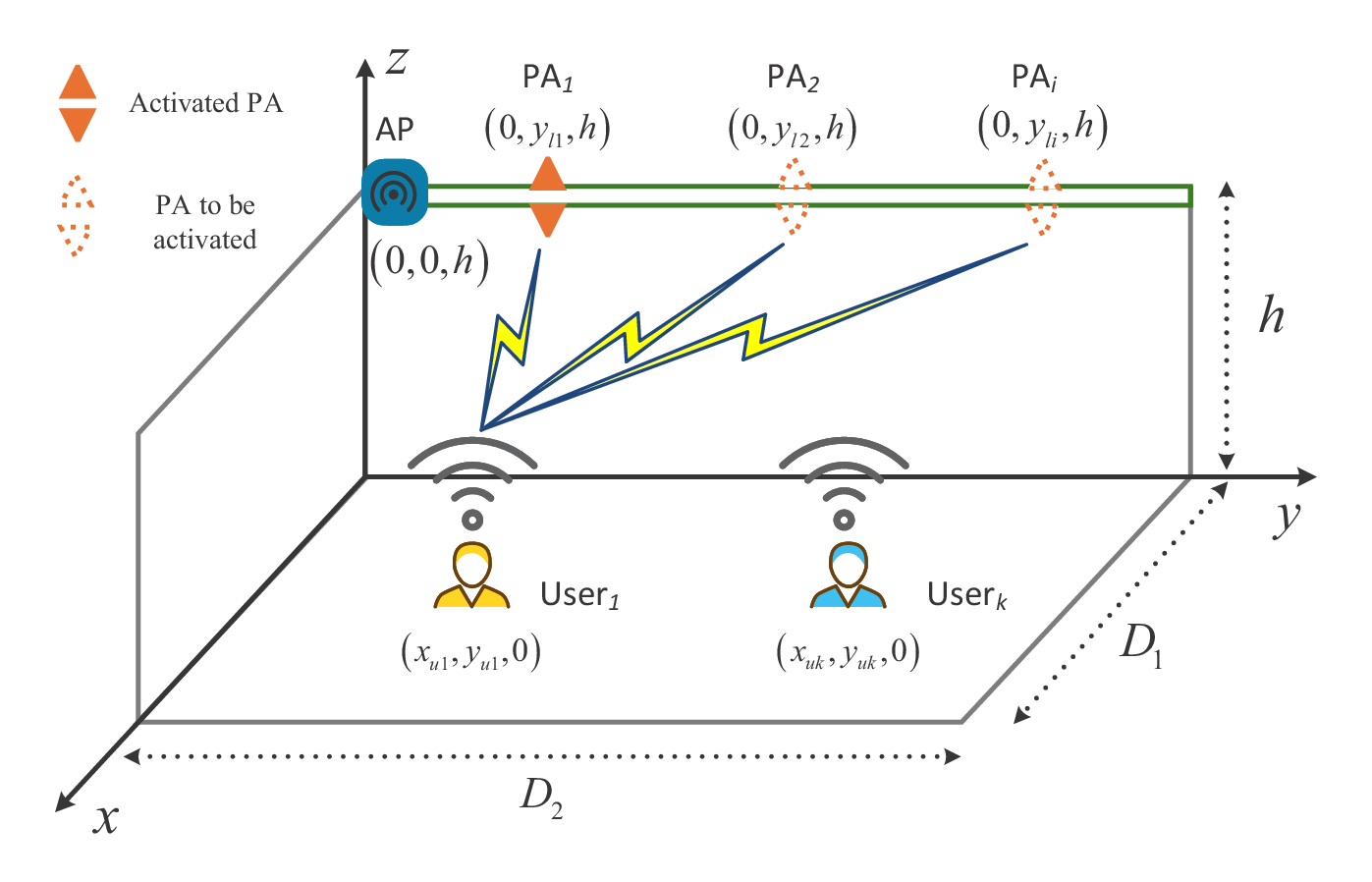}
\caption{PASS-based indoor positioning system model.}
\label{system_model}
\end{figure}

As depicted in Fig.~\ref{system_model}, a signal-waveguide PASS is considered, where $I$ fixed position PAs are deployed to serve $K$ users within a room defined by the spatial dimensions $x \in \left[ {0,D_1} \right]$, $y \in \left[ {0,D_2} \right]$ and $z \in \left[ {0,h} \right]$. An access point (AP) is installed at the ceiling corner of the room, and a waveguide extends along the ceiling boundary, with its span defined as $W \in \left[ {\left( {0,D_2} \right),0,h} \right]$. The location of the $i$-th PA and the $k$-th user are specified by ${L_i} = \left[ 0,{{y_{li}},h} \right]$ and ${U_k} = \left[ {{x_{uk}},{y_{uk}},0} \right]$, respectively, where the locations of the PAs are known, i.e., $x_{li}$ is specified. We assume that the coordinates of AP is $A = \left[ {0,0,h} \right]$.

To facilitate theoretical analysis of fundamental performance bounds and asymptotic characteristics, the large-scale channel is initially characterized under a line-of-sight assumption, which can be modeled as:    
\begin{equation}\label{large-scale channel}
{h_{ls}}\left( {{d_{ik}}} \right) = \frac{c}{{4\pi {f_c}{d_{ik}}}},
\end{equation}
where $c$ represents the speed of light, $f_c$ denotes the carrier frequency, and $d_{ik}$ denotes the distance between the $i$-th PA and the $k$-th user, which can be written as:
\begin{equation}\label{distance between PA and user}
{d_{ik}} = \sqrt {x_{uk}^2 + {{\left( {{y_{li}} - {y_{uk}}} \right)}^2}  + {h^2}}.
\end{equation}

For the small-scale channel model, the channel gain is assumed to be unity through normalization, and we focus only on the phase shift resulting from the propagation distance in free-space conditions. The small-scale channel can be simplified to:
\begin{equation}\label{simplified small-scale fading channel}
{h_{ss}}\left( {{d_{ik}}} \right) = \exp \left( { - \frac{{j2\pi {d_{ik}}}}{\lambda }} \right),
\end{equation}
where $\lambda$ denotes the wave length.

The waveguide used in PASS is a dielectric waveguide, consisting of a bar-shaped dielectric material enclosed by another dielectric medium~\cite{PA2022}. We note that the previous studies have mainly focused on the phase shift caused by waveguide propagation without considering the signal propagation loss in the waveguide. However, the approximation has significant impact on the positioning accuracy, as the neglected attenuation effects directly degrade the ranging accuracy in positioning systems. To address this limitation, it becomes essential to characterize both phase and attenuation parameters through the complex propagation constant, which can be written as~\cite{Microwave_engineering}:
\begin{equation}\label{complex propagation constant}
\gamma  = \alpha  + j\beta  = \sqrt {{k_c} - {\omega ^2}{\mu _0}{\varepsilon _0}{\varepsilon _r}\left( {1 - j\tan \delta } \right)},
\end{equation}
where $\alpha$ and $\beta$ represent the attenuation constant and propagation constant of the waveguide, respectively. ${k_c}$, ${\omega}$ and  ${\mu _0}$ represent the cut-off wave number, angular frequency and vacuum permeability, respectively. ${\varepsilon _0}$, ${\varepsilon _r}$ and $\tan \delta$  denote vacuum permittivity, relative permittivity and loss angle tangent, respectively.

Since most dielectric materials have small losses, i.e., $\tan \delta  \ll 1$, the attenuation constant and propagation constant can be expressed as~\cite{Microwave_engineering}:
\begin{subequations}\label{the attenuation constant and propagation constant}
\begin{align}
&{\alpha} = \frac{{{k_r^2}\tan \delta }}{{2\beta }},\\
&\beta  = \sqrt {{k_r^2} - k_c^2},
\end{align}
\end{subequations}
where $k_r = \omega \sqrt {{\mu _0}{\varepsilon _0}{\varepsilon _r}} $ is the real wave number in the absence of loss. We assume that the real wave number $k_r$ and the cut-off wave number ${k_c}$ satisfy $k_r \gg {k_c}$, then the attenuation constant and propagation constant can be rewritten as:
\begin{subequations}\label{rewritten attenuation constant and propagation constant}
\begin{align}
&\alpha  = \frac{{\pi \sqrt {{\varepsilon _r}} \tan \delta }}{{{\lambda}}},\\
&\beta  = \frac{{2\pi \sqrt {{\varepsilon _r}} }}{{{\lambda}}}.
\end{align}
\end{subequations}

The propagation of electromagnetic waves in the lossy medium can be described as~\cite{Microwave_engineering}:
\begin{equation}\label{propagation of electromagnetic waves in a lossy medium}
l\left( {{y_{li}}} \right) = {e^{ - \gamma {y_{li}}}} = {e^{ - \left( {\alpha  + j\beta } \right){y_{li}}}}.
\end{equation}

\subsection{Signal Model}
We assume that only one PA is activated in each time slot. From a positioning perspective, we only consider uplink signals. The signal of the $k$-th user propagates through free space to the $i$-th PA, and then propagates through the waveguide to the AP. The received signal of the $k$-th user at the AP can be written as:
\begin{equation}\label{The received signal at the AP}
\begin{aligned}
&{r_{ik}} = {h_{ls}}\left( {{d_{ik}}} \right){h_{ss}}\left( {{d_{ik}}} \right)l\left( {{y_{li}}} \right)\sqrt {{P_k}} {s_k} + {n_k}\\
&= \frac{{c{e^{ - \alpha {y_{li}}}}}}{{4\pi {f_c}{d_{ik}}}}\exp \left( { - \frac{{j2\pi \left( {{d_{ik}} + \beta {y_{li}}} \right)}}{\lambda }} \right)\sqrt {{P_k}} {s_k} + {n_k},
\end{aligned}
\end{equation} 
where $P_k$ denotes the transmit power at user $k$, ${n_k}$ is the zero-mean additive white Gaussian noise (AWGN) with the variance ${\sigma ^2}$, and $s_k$ represents the transmitted signal of the $k$-th user.

\section{PASS-based Positioning Approach}

From \eqref{The received signal at the AP}, we can see that for a specific user $k$, the received power of AP propagated through the $i$-th PA can be written as:
\begin{equation}\label{received power}
{P_{ik}} = {\left| {\frac{{c{e^{ - \alpha {y_{li}}}}}}{{4\pi {f_c}{d_{ik}}}}} \right|^2}{P_k} + \chi ,
\end{equation} 
where $\chi \sim \mathcal{N}\left( {0,{\sigma ^2}} \right)$ represents the power of AWGN.

Then, we can utilize RSSI to measure the distance from each PA to the user. The noisy estimated distance between the $i$-th PA and the $k$-th user can be rewritten as:
\begin{equation}\label{the distance}
{\hat d_{ik}} = \frac{{c{e^{ - \alpha {y_{li}}}}}}{{\sqrt {\frac{{{P_{ik}}}}{{{P_k}}}} 4\pi {f_c}}}.
\end{equation}  

By combining (\ref{distance between PA and user}) and (\ref{the distance}), we can obtain the distance measurement equation:
\begin{equation}\label{distance equation}
{\left( {{x_{li}} - {x_{uk}}} \right)^2} + y_{uk}^2 + {h^2} = \frac{{{P_k}}}{{{P_{ik}}}}{\left( {\frac{{c{e^{ - \alpha {y_{li}}}}}}{{4\pi {f_c}}}} \right)^2}.
\end{equation} 

From (\ref{distance equation}), we can see that for each PA, there is a distance corresponding to user $k$. Therefore, we can obtain the equation of a circle centered in $L_i$ with radius $d_{ik}$, which can be expressed as:
\begin{equation}\label{position calculation equation}
\begin{array}{l}
\left\{ {\begin{array}{*{20}{c}}
{x_{uk}^2 + {{\left( {{y_{l1}} - {y_{uk}}} \right)}^2} + {h^2} = \hat d_{1k}^2,}\\
{x_{uk}^2 + {{\left( {{y_{l2}} - {y_{uk}}} \right)}^2} + {h^2} = \hat d_{2k}^2,}\\
 \cdots \\
{x_{uk}^2 + {{\left( {{y_{lI}} - {y_{uk}}} \right)}^2} + {h^2} = \hat d_{Ik}^2.}
\end{array}} \right.\\
\end{array}
\end{equation}

\begin{figure}[t!]
\centering
\includegraphics[width =2.8in]{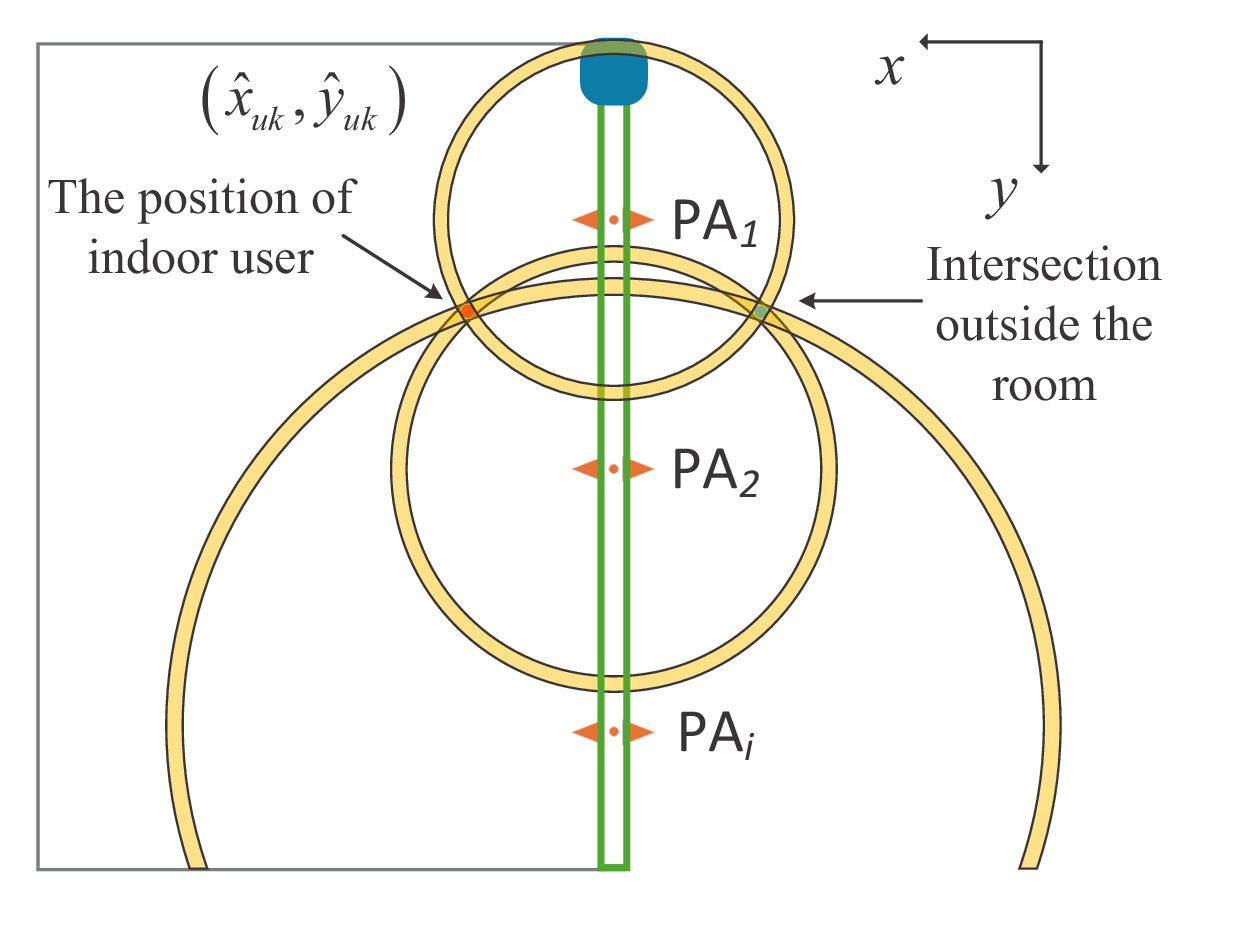}
\caption{Projection map of PASS-based indoor positioning system.}
\label{touying}
\end{figure}

As shown in Fig.~\ref{system_model}, in the signal waveguide scenarios, the center of the circle $L_i$ lies on the same line, so there are two feasible two-dimensional (2D) intersection points in \eqref{position calculation equation}, making it impossible to directly solve using ordinary least squares algorithm. However, we note that since the waveguide is distributed at the edge of the ceiling, one of the intersection points is outside the wall, which is shown in Fig.~\ref{touying}. We can exploit this characteristic to design an appropriate solution algorithm.

\begin{algorithm}[!ht]
    \renewcommand{\algorithmicrequire}{\textbf{Input:}}
	\renewcommand{\algorithmicensure}{\textbf{Output:}}
	\caption{PASS-based WLS Algorithm}
    \label{single-waveguide}
    \begin{algorithmic}[1] % 控制是否有序号
        \REQUIRE  The coordinate of each PA $\left[ 0,{{y_{li}},h} \right]$, the received signal power $P_{ik}$, the estimated distance $\hat{d}_{ik}$ and the noise power $N_i$. % input 的内容
	    \ENSURE The coordinate of user $\left[ {{\hat{x}_{uk}},{\hat{y}_{uk}},0} \right]$. % output 的内容
        
        \STATE Define ${b_i} = \hat d_{ik}^2 - y_{li}^2 - {h^2}$ and $v = x_{uk}^2 + y_{uk}^2$.
        \STATE According to \eqref{rewritten position calculation equation}, construct the matrices ${\bf{A}}$, ${\bf{b}}$, ${\bf{Y}}$. 
        \STATE  Construct the weight matrix based on SNRs: \\${\bf{w}} = {\rm{diag}}\left( {\frac{{{P_{1k}}}}{{{N_1}}},\frac{{{P_{2k}}}}{{{N_2}}}, \cdots ,\frac{{{P_{Ik}}}}{{{N_I}}}} \right)$.
        \STATE  The solution ${\bf{\hat Y}} = {\left( {{{\bf{A}}^T}{\bf{wA}}} \right)^{ - 1}}{{\bf{A}}^T}{\bf{wb}}$.
        \STATE  Calculate the coordinate of y-axis: ${\hat x_{uk}} = \sqrt {\hat v - \hat y_{uk}^2}$.
        \STATE \textbf{return} $\hat{x}_{uk}$, $\hat{y}_{uk}$.
    \end{algorithmic}
\end{algorithm}

As summarized in algorithm~\ref{single-waveguide}, we firstly define ${b_i} = \hat d_{ik}^2 - y_{li}^2 - {h^2}$ and $v = x_{uk}^2 + y_{uk}^2$. Then, \eqref{position calculation equation} can be rewritten as:
\begin{equation}\label{rewritten position calculation equation}
\left\{ {\begin{array}{*{20}{c}}
{ - 2{y_{l1}}{y_{uk}} + v = {b_1}},\\
{ - 2{y_{l2}}{y_{uk}} + v = {b_2}},\\
 \cdots \\
{ - 2{y_{lI}}{y_{uk}} + v = {b_I}}.
\end{array}} \right.
\end{equation}

To simplify the presentation and facilitate calculations, \eqref{rewritten position calculation equation} can be rewritten into a matrix form as:
\begin{equation}\label{matrix form position calculation equation}
{\bf{AY}} = {\bf{b}},
\end{equation}
where 
\begin{align*}
&{\bf{A}} = \left[ {\begin{array}{*{20}{c}}
{ - 2{y_{l1}}}&1\\
{ - 2{y_{l2}}}&1\\
 \vdots & \vdots \\
{ - 2{y_{lI}}}&1
\end{array}} \right],\\
&{\bf{Y}} = {\left[ {\begin{array}{*{20}{c}}
{{y_{uk}}}&v
\end{array}} \right]^T},\\
&{\bf{b}} = {\left[ {\begin{array}{*{20}{c}}
{{b_1}}&{{b_2}}& \ldots &{{b_I}}
\end{array}} \right]^T}.
\end{align*}

We can see that there are two unknowns parameters in \eqref{rewritten position calculation equation}, therefore, the equation is solvable when at least two PAs are present. Since the distance from the user to each PA varies, the signal-to-noise ratios (SNRs) at each PA differ, leading to varying levels of reliability in the distance measurements. Thus, we define the weight matrix as:
\begin{equation}\label{weight matrix}
{\bf{w}} = {\rm{diag}}\left( {\frac{{{P_{1k}}}}{{{N_1}}},\frac{{{P_{2k}}}}{{{N_2}}}, \cdots ,\frac{{{P_{Ik}}}}{{{N_I}}}} \right),
\end{equation}
where ${N_i}$ denotes the noise power, $i = 1,2, \cdots ,I$.

The WLS solution to \eqref{matrix form position calculation equation} is given by:
\begin{equation}\label{WLS solution}
{\bf{\hat Y}} = {\left( {{{\bf{A}}^T}{\bf{wA}}} \right)^{ - 1}}{{\bf{A}}^T}{\bf{wb}}.
\end{equation}

There are two solutions for the $y$-coordinate of the user. We take the $y$-coordinate value distributed inside the room, which can be written as:
\begin{equation}\label{y_uk}
{\hat x_{uk}} = \sqrt {\hat v - \hat y_{uk}^2}.
\end{equation}

\begin{remark}\label{remark1:I on positioning}
When the number of PAs $I$ is greater than the number of unknowns, \eqref{matrix form position calculation equation} becomes an overdetermined equation, and WLS algorithm solves $\bf Y$ by minimizing the weighted sum of squared residuals. As the number of PAs increases, the number of independent observations increases, allowing the random errors in the measurement vector $\bf b$ to more effectively offset each other during the solution process. Therefore, as the number of PAs increases, the positioning error decreases.
\end{remark}

\begin{remark}\label{remark2:I on robustness}
As the number of PAs increases, the random errors of the observation vector $\bf b$ in \eqref{matrix form position calculation equation} are statistically averaged during the solution process, reducing the sensitivity of the solution to any single noisy measurement, thereby enhancing the robustness of the system.
\end{remark}

\section{Simulation Results}

Here, we present numerical results regarding the performance evaluation of PASS-based WLS algorithm. $I$ PAs are evenly distributed on the waveguide at the edge of the ceiling, and only one PA is active in a time slot. We conduct simulations to evaluate the positioning accuracy by considering the impact of the number of PAs and user location distribution, and then analyze the positioning robustness of different numbers of PAs. The spatial dimensions of the room are defined by $x \in \left[ {0,6} \right]$, $y \in \left[ {0,10} \right]$ and $z \in \left[ {0,3} \right]$. We have set the carrier frequency to 2.8 GHz and the transmission power of the user to 0.1 W. The relative permittivity and loss angle tangent are set to ${\varepsilon _r = 2.08}$ and $\tan \delta = 0.0004$, respectively~\cite{Microwave_engineering}. 

\begin{figure}[t!]
\centering
\includegraphics[width =3.2in]{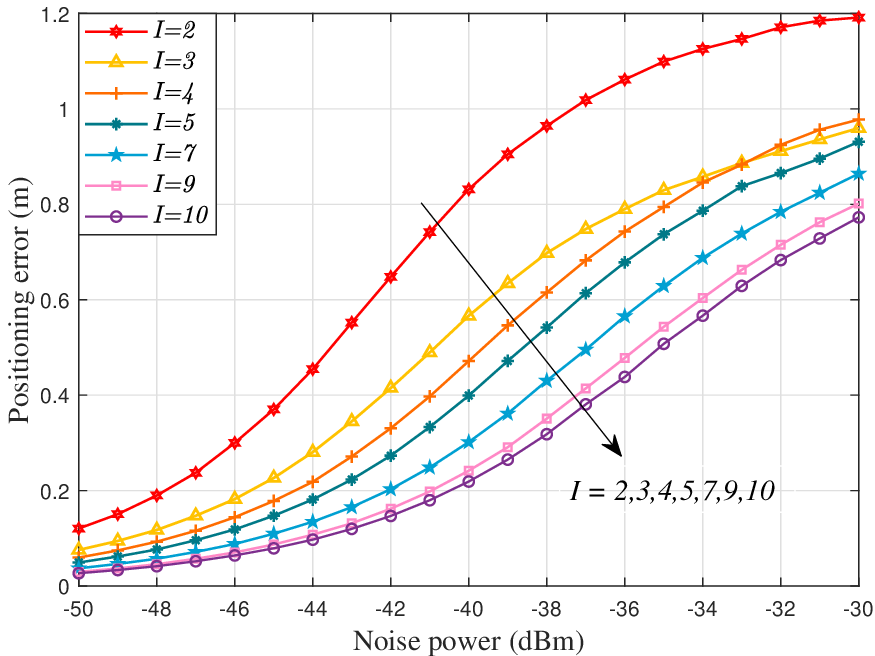}
\caption{Positioning error under different numbers of PAs.}
\label{single-waveguide position Error}
\end{figure}

\emph{1) Impact of the numbers of PAs:} Fig.~\ref{single-waveguide position Error} depicts the positioning error versus noise power for different numbers of PAs. We can see that as the noise power increases, the positioning error also increases for all number of PAs, which suggests that the high noise power negatively affect the positioning accuracy. As the number of PAs increases, the positioning error gradually decreases, indicating that the more PAs leads to the higher the positioning accuracy, which validates our~\textbf{Remark~\ref{remark1:I on positioning}}. Nonetheless, the impact of the number of PAs on positioning error is not linear. The reduction in positioning error is more significant from $I$ = 2 to $I$ = 5. However, as the number of PAs continues to increase, e.g., from $I$ = 7 to $I$ = 10, the reduction in positioning error becomes slower. This suggests that when the number of PAs is low, increasing the number of PAs can significantly improve positioning accuracy, while at higher numbers of PAs, further increases lead to diminishing improvements in positioning accuracy.

\begin{figure}[t!]
\centering
\includegraphics[width =3.5in]{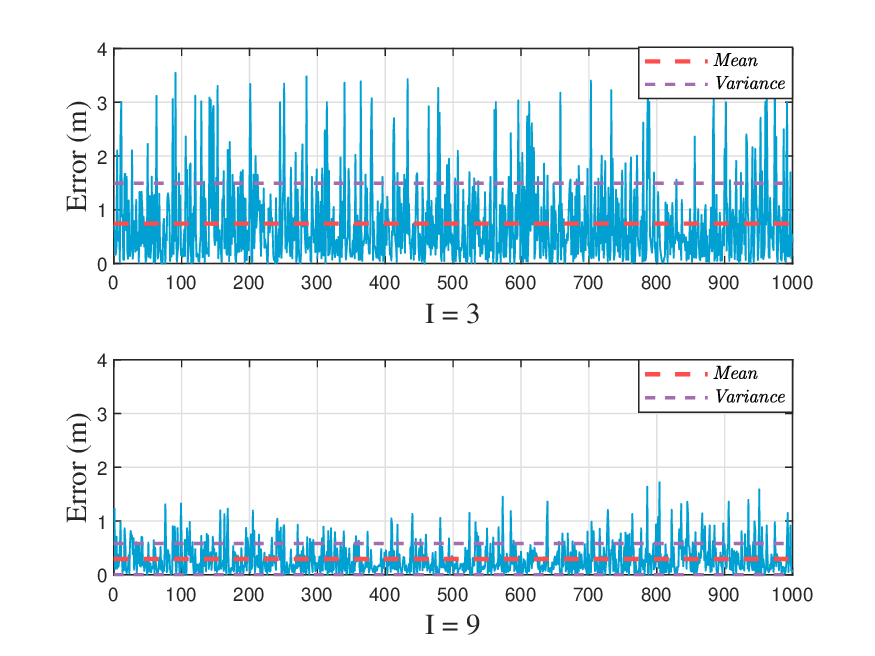}
\caption{Mean and variance of the positioning error under different numbers of PAs. ${\sigma ^2}$ = -40 dBm.}
\label{rubust_test}
\end{figure}

\emph{2) The positioning robustness of different numbers of PAs:} The results from 1000 Monte Carlo simulations are illustrated in Fig.~\ref{rubust_test}. The blue lines represent the error in each trial, while the red dashed line indicates the mean error, and the purple dashed line represents the variance. Both plots show significant fluctuations in the error values across the different simulations, indicating the impact of random noise on the positioning error. When $I$ = 3, the mean error and variance are 0.74 m and 0.75, respectively, whereas for 
$I$ = 9, the mean error and variance decrease to 0.30 m and 0.29, indicating improved positioning accuracy and robustness with more PAs, which validates our~\textbf{Remark~\ref{remark2:I on robustness}}. This supports the earlier finding that higher number of PAs can generally improve positioning accuracy.     

\begin{figure}[t!]
\centering
\includegraphics[width =3.5in]{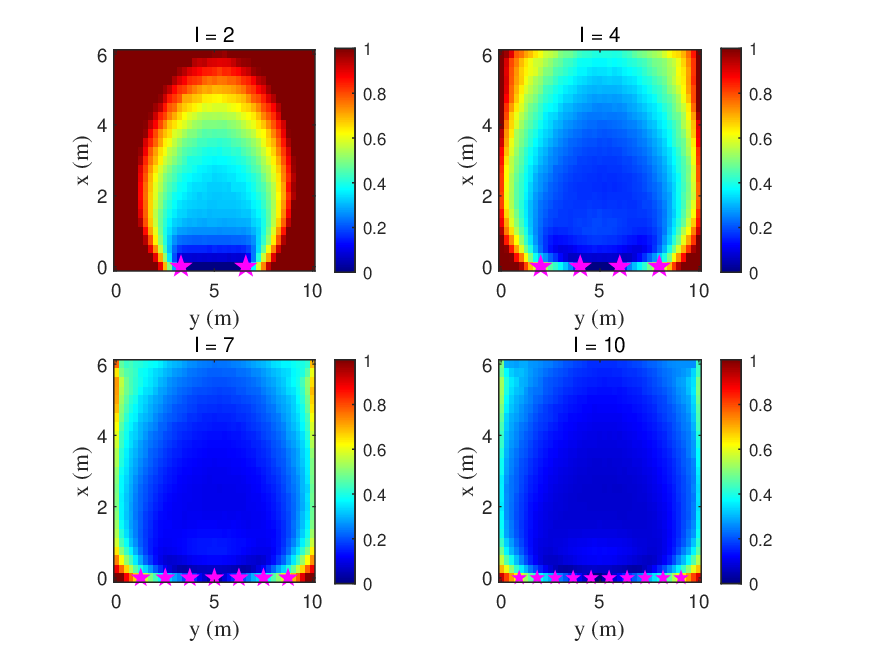}
\caption{User positioning error map under different numbers of PAs. ${\sigma ^2}$ = -40 dBm.}
\label{differ_position_differ_PA}
\end{figure}

\emph{3) The impact of user distribution on positioning accuracy:} Fig.~\ref{differ_position_differ_PA} shows the heat map of spatial positioning error under different numbers of PAs. The x-axis and y-axis are spatial coordinates, and the color bar represents the normalized positioning error. Blue represents a smaller error and red represents a larger error. The purple asterisk marks the position of the PAs. We can see that the positioning error is small when the user is distributed among PAs, while the positioning accuracy greatly decreases when the user is distributed outside of PA. The color blocks gradually change when the user is far away from the PAs, and the positioning accuracy decreases. We can also observe that as the number of PAs increases, the red covered area decreases and the blue covered area increases, which indicates that as the number of PAs increases, the area with large positioning errors is significantly reduced, and the system can achieve low-error positioning in more areas.

\section{Conclusions}

In this letter, we started with a review of earlier work on flexible antennas, and we then proposed the novel PASS-based indoor positioning models. To evaluate the position of user, we presented the PASS-based WLS algorithm. Simulations were performed to verify the correctness and robustness of the algorithms. Future work will focus on extending these models to more complex environments, including non-line-of-sight conditions and dynamic user movement. Additionally, optimizing the antenna configuration and incorporating machine learning techniques for enhanced positioning accuracy and adaptability represent promising directions.

\bibliographystyle{IEEEtran}
\bibliography{IEEEabrv,PASS-based_indoor_position}

\end{document}